# Detection of organic materials by spectrometric radiography method

S. V. Naydenov*, V. D. Ryzhikov, G. M. Onyshchenko, P. Lecoq, C. F. Smith

*Abstract* —In this paper we report a spectrometric approach to dual-energy digital radiography that has been developed and applied to identify specific organic substances and discern small differences in their effective atomic number. An experimental setup has been designed, and a theoretical description proposed based on the experimental results obtained. The proposed method is based on application of special reference samples made of materials with different effective atomic number and thickness, parameters known to affect X-ray attenuation in the low-energy range. The results obtained can be used in the development of a new generation of multi-energy customs or medical X-ray scanners.

## I. INTRODUCTION

Presently, most dual-energy X-ray security systems for luggage inspection as well as for medical diagnostics use multi-detector systems operating in the current mode. Such X-ray instruments rely on two types of detectors – one a "thin" scintillator with low effective atomic number which therefore transmits the high energy portion of the incident radiation, i.e., a low energy detector (LED); and the second a "thick" scintillator with high effective atomic number seeking near total absorption of the incident radiation, i.e., a high energy detector (HED) (see, for example, [1-5]). Such combined detector systems can receive and analyze two type of signals: one from the LED detector, with its response to both the low-energy and the high-energy components; and the other from the total absorption (HED) detector, which records predominantly the high-energy component.

The dual-energy approach allows substantial improvement of the image quality of X-ray inspection systems [6-13] as compared with systems based on conventional broad spectrum X-ray radiography. However, in this method there is an unavoidable source of error due to the presence in the LED detector of a large fraction of high-energy quanta which interfere with the detection of the low-energy quanta. This ultimately affects the quality of the obtained image, especially in the identification of organic materials with small values of $Z_{eff} \propto 5-10$.

This drawback can be largely avoided by fabricating the LED with specially designed scintillators based on doped zinc selenide compounds such as ZnSe(Te), which have a low $Z_{eff} \approx 33$ and a low density $\rho = 5.4\,g/cm^2$. With such scintillators and rather small detector thickness (e.g., hundreds of microns), absorption in the high energy range (80-90 keV) does not exceed 10-15%, i.e., the detector shows a roughly spectrometric behavior. A unique combination of useful physical characteristics, accompanied by high light output (up to 120-140% in comparison to CsI(Tl)), allow its application in inspection equipment as the most efficient scintillator (among the known variety of scintillation materials) for low energy detectors. Its use in customs inspection systems enable high-speed automatic sorting of loads with potentially dangerous enclosures [1]. In applications of this scintillator for medical diagnostics (i.e., the study of biological objects in 3D tomographic mode), it is possible not only to observe a clear difference between muscular and osseous tissues, but to reliably detect small deviations (2-3%) of calcium content in the bone tissue [3].

It has been confirmed [4] that the use of ZnSe(Te) in the LED component of a dual energy system ensures detection with high probability of illegal and dangerous substances in loads and luggage. At the same time, the engineering solution proposed in that patent and the specified design of the dual-energy detector array do not allow detection of all potentially dangerous substances with sufficiently low error probability. However, a theoretical approach developed to exploit the potential of multi-monochromatic radiation in several narrow and mutually separated spectral ranges [5] predicts very high (90-95%) accuracy in determination of $Z_{eff}$ and other important parameters of the inspected object. Such accuracy would allow detection of many materials with nearly 100% probability, e.g., plastic explosives in solid or liquid form, during security inspection. In medical diagnostics, this would allow early recognition of tumor formation, osteoporosis, cardiovascular atherosclerosis, etc.

This work was supported in part by NATO SfP-982823 and STCU 4115 projects.

*Corresponding author: S.V. Naydenov is with Concern "Institute for Single Crystals", Kharkov, Ukraine. Phone: +38-057-341-03-51, fax: +38-057-719-59-97; e-mail: sergei.naydenov@gmail.com

V. D. Ryzhikov and G. M. Onyshchenko are with the Institute for Scintillation Materials of the National Academy of Sciences of Ukraine.

P. Lecoq is with European Research Centre, European Organization for Nuclear Research (CERN), Geneva, Switzerland. E-mail: paul.lecoq@cern.ch

C.F. Smith is with Lawrence Livermore National Laboratory (LLNL), Livermore, CA, USA. E-mail: cfsmith@nps.edu

## II. EXPERIMENTAL SETUP AND THEORY

In this paper, we aimed at further development of the dual-energy approach in digital radiography by exploiting the advantages for determination of material parameters offered by the spectrometric method. Obviously, the spectrometric method, as distinct with the current method, provides a way to solve the energy separation problem, since this method allows detection of signals at essentially any emission energy from the continuous spectrum of an X-ray tube, ensuring rather high energy resolution. This removes certain problems related to insufficient energy separation in traditional dual-energy radiography leading to lower accuracy. Therefore, in spite of a certain level of added technical complication, the spectrometric approach seems very promising from the viewpoint of multi-energy radiography. Positive results in this experimental work would not only confirm the correctness of the overall spectrometric approach, but it would also provide insight as to the best technical approach to be taken in its implementation, e.g., the use of special multi-monochromatic filters or other sources of monochromatic radiation for improvement of inspection accuracy. In our development of this spectrometric approach, we have designed an experimental installation to carry out a set of experiments to detect and identify different organic materials by measuring their effective atomic number and constructing a theory basis for interpreting the experimental facts.

The experimental setup is shown in Fig. 1. The flux of quanta from X-ray tube 1 passes through collimator 2, inspected sample 3 and detector collimator 9. The resulting X-ray signal was recorded by the detector system consisting of PMT 4 and scintillation crystal 5. The valid signal from the PMT anode was amplified by linear spectrometric amplifier 6, processed by analog-to-digit converter 7 and stored in the computer 8. As a source of X-ray radiation, we used an X-ray tube operating at 120 kV, with anode current 15 μA. Working loads of the spectrometric circuit did not exceed 3000 s$^{-1}$.

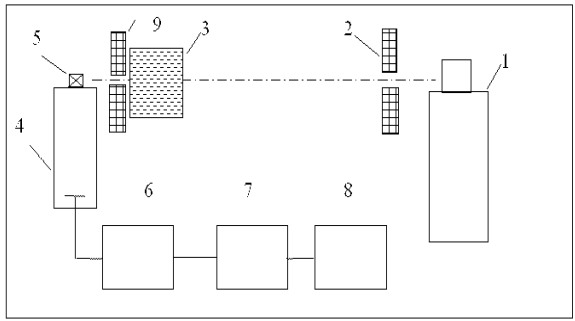

Fig. 1. Experimental set-up. 1 – X-ray tube; 2 and 9 – lead collimators; 3 – inspected reference sample; 4 – PMT; 5 – scintillation crystal ZnSe(O, Te); 6 – linear spectrometric amplifier; 7 – ADC; 8 – pulse analyzer.

recorded by this spectrometer, together with a calibration spectrum formed from the emissions from Am-241, Cs-137 and Co-57 sources. Both the X-ray tube and the calibration spectra were obtained under the same conditions. As a detector, we used a ZnSe(O, Te) scintillator of dimension 3x3x0.5 mm coupled to a Hamamatsu R1306 PMT. The shaping time of the valid signal was ~ 5 μs. Note that photons of ~ 73 keV energy are also formed as a result of excitation by quanta of the lead collimator.

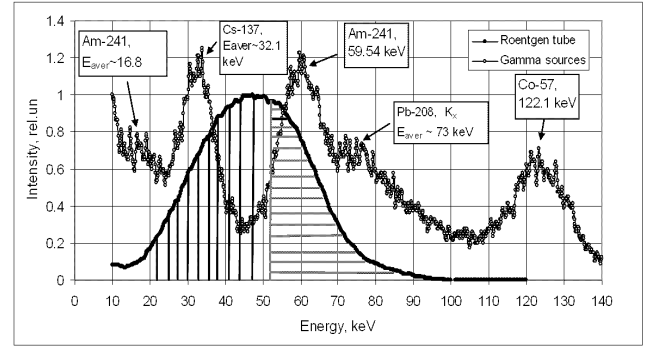

Fig. 2. Measured energy spectra of photons from X-ray tube and radioactive sources Am-241, Cs-137, Co-57.

To process the signal, the photons that passed through the inspected samples were separated into two energy ranges: ~20-50 keV and 50-100 keV. Fig.3 shows the characteristic distortions of an X-ray radiation spectrum after passing through materials that differ slightly in $Z_{eff}$ and density. Each line on this figure is a sort of "fingerprint" for the respective material.

In Table I, values are listed of effective atomic number of "light" materials from paraffin to aluminum, calculated to account for the photo effect as the main process of interaction of quanta with the inspected material in the energy range 20–100 keV.

TABLE I. EFFECTIVE ATOMIC NUMBER OF SOME "LIGHT" SUBSTANCES

| Material | $Z_{eff}$ |
|---|---|
| Paraffin | 5.42 |
| Polystyrene | 5.70 |
| Scintillator based on polystyrene | 5.72 |
| Carbon (graphite) | 6.00 |
| Polymethyl metacrylate | 6.48 |
| Soap | 6.84 |
| Glycerol | 6.87 |
| Water | 7.43 |
| Air | 7.65 |
| Teflon | 8.44 |
| Ebonite | 11.54 |
| Quartz glass | 11.58 |
| Aluminum | 13.00 |
| Window glass | 13.15 |

The energy spectrum shown in Fig. 3 confirms that it is possible in principle to obtain the values of a particular measure, parameter *P* (detailed later in this paper) that is sensitive to $Z_{eff}$, for experimental samples of aluminum, Teflon and polymethyl metacrylate prepared according to the above requirements.

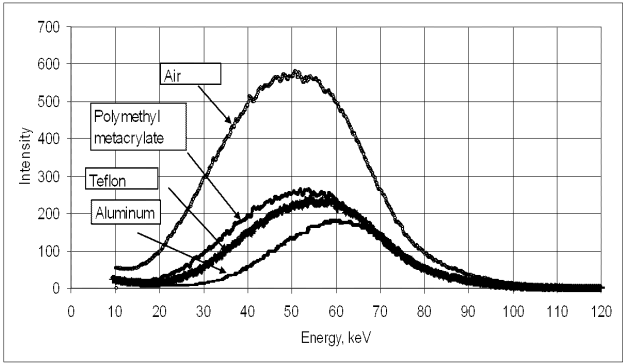

Fig. 3. Experimental energy spectra measured after X-ray tube radiation passed through samples: 34.7 mm polymethyl metacrylate (plexiglass), 18.4 mm Teflon, 9.8 mm aluminum.

Let us describe the spectrometric approach within the framework of a simple theoretical model. Radiation is attenuated in an object according to an exponential law with a linear attenuation coefficient $\mu(E, Z, \rho)$, which depends upon the radiation energy $E$ and the atomic composition, i.e., the effective atomic number $Z$ and density $\rho$ of the material. The radiation is attenuated not only inside the object of thickness $l$, but also inside the scintillator of thickness $l_s$. In a general case, attenuation can also occur in the filter separating the LED and HED assemblies.

In principle, two ways (modes) are possible for detection of this attenuated radiation – a more common *current mode* and *spectrometric mode*. Thus, in our case of dual-energy radiography both detectors (LED and HED) operate in the current mode when they record a signal that is proportional to the photocurrent in the photodiode (which is caused by the internal photo effect due to scintillations in the scintillator). The obtained electric signal is amplified and digitized in the receiving-detecting circuit. The magnitude of the photocurrent (signal) corresponds to a certain averaged intensity value of the radiation passed through the inspected object and attenuated in a specified energy range – in that low- or high-energy range where the radiation is recorded by the corresponding LED and HED detectors with the best efficiency. In the spectrometric mode, the same dual-energy radiography requires only one scintillation detector, which, however, should operate itself in the spectrometric mode. Such detector is coupled to a spectrometric receiving-detecting circuit and can record the attenuated radiation at every energy value with a certain step corresponding to the full energy resolution of the detector. This detector can also record an integral signal in a certain energy range, e.g., in the low- or high-energy range, respectively. Thus, in the above sense one detector operating in the spectrometric mode can, in fact, replace several (two or more, in the case of multi-energy radiography) detectors operating in the current mode. Now, let us consider radiation detection in the spectrometric mode.

Accounting for continuous spectrum of the X-ray source and introducing an instrument function of the detecting system $f(E)$, which depends not only on the radiation spectrum, but also on the detector properties, the full recorded signal $I$ can be written in the integral form

$$I = \int I(E)dE = \int_{\{L,R\}} \exp\left[-\mu_m(E,Z)\rho l\right] f(E) dE, \quad (1)$$

where the integration is over all the energy spectrum, conventionally divided into two ranges ("left window" and "right window"). In Eq. (1), these energy ranges are denoted by letters $L$ and $R$. Also, the expression (1) involves a mass attenuation coefficient $\mu_m$, which depends not on bulk physical properties, but on the electron density of the material and is proportional to the absorption (scattering) cross-section. The integral character of (1) does not determine, in the general case, any quantitative dependencies. However, under the assumption of weak absorption inside the object, such dependencies can be easily found. The assumption of weak absorption corresponds to the mathematical condition $\mu l \ll 1$, which is valid for objects of sufficiently small thickness or materials with low atomic number. In this case, the exponent in the integrand of Eq. (1) can be expanded in series over powers of the said small parameter, retaining only the main contribution terms. After simple transformations, we obtain the relationship

$$I_0 - I = \left(aZ^{q+1} + bZ\right)\rho l, \quad (2)$$

derived by recognizing that the attenuation coefficient is in the form of two summands that are responsible for the two principal absorption mechanisms (Compton and photo-effect scattering) in the energy range characteristic for X-ray emitters. These terms describe absorption due to the photo effect, which is proportional to $Z^{q+1}$ (where $q$ usually can vary from 2 to 4) and absorption due to the Compton effect, with linear dependence on $Z$. The constants $a$ and $b$ include the energy dependence of said absorption cross-sections integrated over all the absorption spectrum. $I_0$ is the recorded integral signal in the absence of the object, which corresponds to the measurement background.

Let us split the full signal $I = I_L + I_R$ into two components $I_L$ and $I_R$, which can be conditionally considered as corresponding to the low- and high-energy range of the spectrum. Now, instead of the pair of constants $\{a,b\}$, we deal with four values $\{a_L, b_L\}$ and $\{a_R, b_R\}$, which correspond to integration of the $I(E)$ signal over the "left" and "right" window of the spectrum, respectively. We assume for simplicity that the "left" window is dominated by the photo effect, and the "right" window – by the Compton scattering effect; this is quite possible for some materials in a certain energy range. In this case $a \approx a_L \gg a_R$ and $b \approx b_R \gg b_L$. Then let us introduce the reduced components $\Delta I_L = I_{0L} - I_L = a_L Z^{q+1} \rho l$ and $\Delta I_R = I_{0R} - I_R = b_R Z \rho l$ for the full signal, with $I_0 = I_{0L} + I_{0R}$ for the background signal. Let a new parameter $P$ be, by definition, the ratio of the partial signal components

$$P = P(Z) = \frac{\Delta I_L}{\Delta I_R}. \quad (3)$$

It is obvious that within the framework of the proposed model this parameter depends only on effective atomic number $Z$ and does not depend upon density or thickness of the material. From equation (3) we obtain

$$P(Z) \propto Z^q. \quad (4)$$

In going from a material with effective atomic number $Z_1$ to another material with effective atomic number $Z_2$, the respective detection parameters $P_{1,2}$ will be related as

$$q = \frac{\ln(P_1/P_2)}{\ln(Z_1/Z_2)}. \quad (5)$$

We assume that the "right" component of $I_R$ is the same for a set of studied objects, i.e., for any pair of such objects $I_{1R} = I_{2R}$ or $\Delta I_{1R} = \Delta I_{2R}$, since the background signal is the same for all objects. In practice, this can be achieved by varying the material thickness or by additional calibration of the whole signal (including the "right" component), which would level the experimental $I_R$ for different objects. Then we may consider $P$ as proportional to the low-energy "left" component of the signal, $P \propto \Delta I_L$. In this case Eq. (5) can be written as

$$q = \frac{\ln(\Delta I_{1L}/\Delta I_{2L})}{\ln(Z_1/Z_2)} \approx \text{const} \quad (6)$$

This relationship is frequently observed when reviewing experimental data. Thus, passing from paraffin ($Z = 5.42$) to water ($Z = 7.43$) corresponds to the change in $P$ (directly related to the measured values) from $P = 5.45$ to $P = 11.87$. This results in $q = 2.47$, which differs from the theoretical value because of the simplified character of the proposed model, which only roughly accounts for different character of absorption for radiation of different energy. However, of essential importance is the fact that there is a clear one-to-one correspondence (under certain experimental conditions) between the effective atomic number of the material and its spectrometric response to X-ray inspection in the multi-energy mode. We have studied this relationship experimentally for organic materials.

One should also note the following. Equations (2) and (4) suggest the linear relationship

$$\ln P = A \ln Z_{\text{eff}} + B, \quad (7)$$

where $A$ and $B$ are certain constants. It can be easily shown that this is true upon condition

$$\frac{b_L}{b_R} \ll P \ll \frac{a_L}{a_R}, \quad (8)$$

which indicates which absorption mechanism is predominant and to what extent each of the spectral windows prevails. One can formally generalize (8) for the case when no assumptions on weak absorption in the object are made. Then this relationship will take the form of the following law

$$\varphi(\ln(I_L/I_R)) = A \ln Z_{\text{eff}} + B \quad (9)$$

with constant parameters $A$ and $B$, as well as a certain one-valued function $\varphi(...)$. In deriving of Eq. (9), we, in fact, replaced the logarithm of the mean value by the mean value of the logarithm, i.e., assumed that $\ln(\langle...\rangle) \approx \langle\ln(...)\rangle$, where $\langle...\rangle$ means integration over the energy spectrum with the appropriate weight (instrument function or single response function). This approximation is justified in the case of weak absorption. With a certain error, it can be assumed in the case of arbitrary absorption. Then we can use (9) as a theoretical foundation of the experimentally observed one-to-one relationship between the effective atomic number and the logarithm of the response ratio in the "left" and 'right" spectral window in dual-energy radiography. It is natural to use this feature for material recognition. A theoretical accuracy limit of such spectrometric method follows from Eq. (7). Error $\Delta Z$ in determination of $Z$ is related to the recording error $\Delta P$ in spectrometric measurement of parameter $P$:

$$\left(\frac{\Delta Z}{Z}\right) \propto \left(\frac{1}{q}\right)\left(\frac{\Delta P}{P}\right), \quad (10)$$

since one can assume $A \approx q$, with $q$ taking on characteristic values from 2 to 3 in the low-energy region for organic materials with relatively small effective atomic number. It is clear that improving the signal recording accuracy one can achieve high accuracy in determination of $Z_{\text{eff}}$. The accuracy is higher with larger $q$ or $P$.

### III. DISCUSSION OF RESULTS

Thus we have made an attempt to improve the quality of identification of organic materials using a spectrometric procedure for the detection of X-ray quanta passing through an inspected object. Comparative analysis of the current and spectrometric methods has been carried out. The proposed procedure is based on the use of special reference samples. Using these samples, an array of values of a certain parameter $P_{\exp}$, which are proportional to the degree of attenuation of the quanta flux by these reference samples in the low-energy range, is obtained. In other words, $P_{\exp}$ is proportional to the area under the corresponding curve in Fig. 3 (in a certain radiation energy range). This experimental parameter corresponds to the theoretical parameter introduced above (3) upon the condition that neither of the values $I_L$ or $I_R$ is fixed. In our experiments, we chose the sample thickness in such a way that the value of $I_R$ remained fixed. In this case $P = P_{\exp}$ and

$$P_{\exp} \propto I_L \equiv \int_{\{L\}} I(E) dE; \quad I_R = \text{const}, \quad (11)$$

where value $I(E)$ denotes the recorded spectral signal at the spectrometer output for a specified radiation energy, and $\{L\}$ is the "left" window range of full energy spectrum.

The proposed procedure uses specially prepared reference samples of pre-calculated thickness made of materials with different $Z_{eff}$. These samples are required for the preliminary calibration of the measurement circuit. A data array is calculated for intensity values of X-ray radiation transmitted through these reference samples. In this data array, each $P_i$ value is a function of $(Z_{eff})_i$ of the samples in the low-energy range. The obtained array of parameters $P_i$ can be used, e.g., for preliminary calibration of the color palette of the observation monitors.

In the proposed procedure the thickness of reference samples is chosen from the condition of equal absorption probability for each of the studied set of substances in the high-energy range 50-100 keV. This can be clearly seen in Fig. 3, where the expected effect (dependence of the spectrometric response from the type of material) is very significant in the low-energy range of 20-50 keV. A small error persists due to the shift of the maximum of the obtained energy spectrum by ~ 10 keV towards higher energies.

For comparative analysis of the traditional and proposed, i.e. current and spectrometric methods, the parameter $P = I_L + I_R$ in modeling of the current mode was assumed proportional to the total signal from the thin (LED) and thick (HED) detectors. The sample thickness was chosen from the requirement of equality of attenuation coefficient values for a given set of substances in the high-energy range 50-100 keV (Fig. 4). Substantial deviation can be noted from a simple dependence of $P$ as function of $Z_{eff}$ in this case.

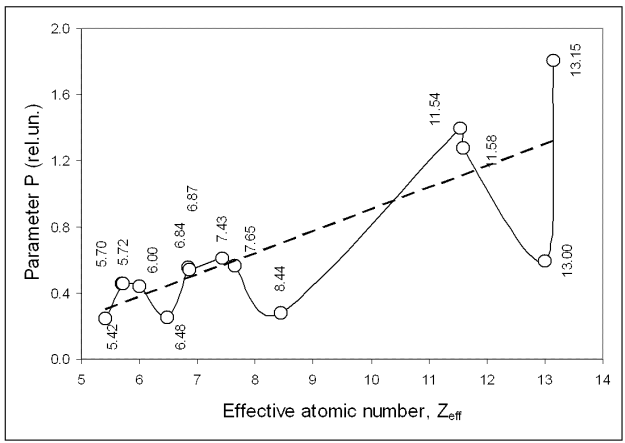

Fig. 4. Dependence of parameter $P \propto (I_L + I_R)$ proportional to the total signal from low-energy and detectors as function of $Z_{eff}$.

It is different if, in the spectrometric approach, one uses as parameter $P$ a function proportional to the signal obtained from only the low-energy region (e.g., at ~20-50 keV), and attenuation is the same for all the samples in the high-energy region of the working range (at ~50-100 keV in our case), the theoretical plot becomes a one-to-one correspondence dependent on the parameter $P \propto I_L$ on $Z_{eff}$ (Fig. 5). In other words, each specified value of $P$ corresponds to a unique value of $Z_{eff}$, and vice versa. All statistical criteria for approximation of experimental data are satisfied. Moreover, this dependence remains basically unchanged with different thickness of the inspected samples.

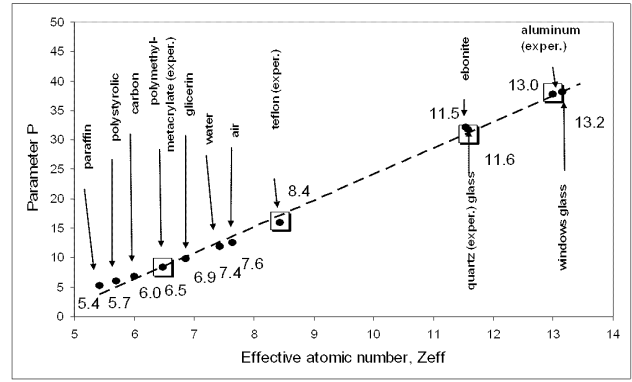

Fig. 5. Theoretical (dots) and experimental (rectangles) dependences of another parameter $P \propto I_L$ ($I_R = \text{const}$ for all references), proportional to the signal only from the "thin" (low-energy) detector, as function of effective atomic number $Z_{eff}$ for a set of "light" substances.

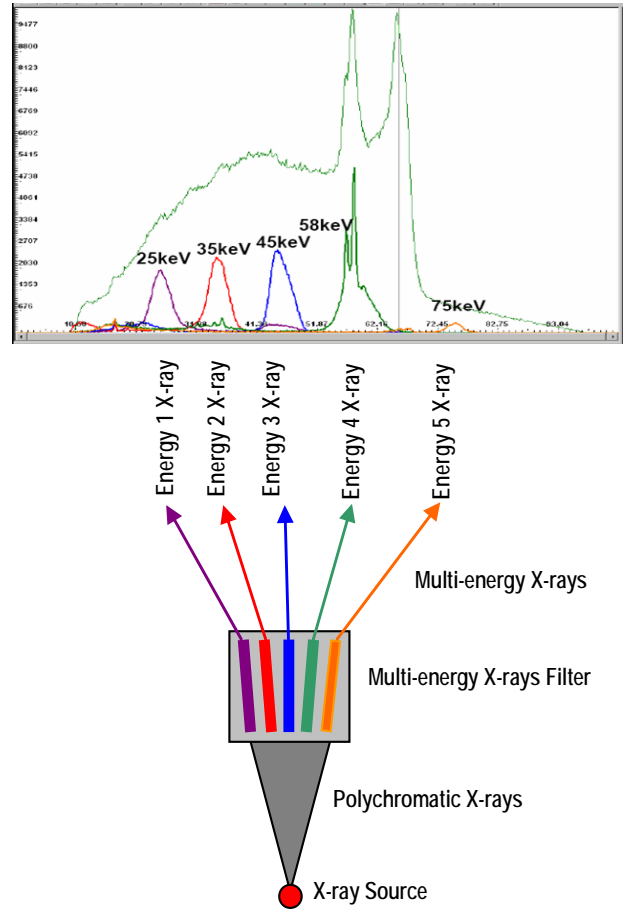

Fig. 6. A scheme of a Multi-energy X-ray Filter with each coated mirror designed (Bragg's Law) for a different X-ray energy.

It should be noted that, though in our experiments the spectrometric method demonstrated its high accuracy, its practical use remains difficult because it requires rather complicated equipment, and it is not possible to carry out measurements and visualization of radiographic images on the real time scale. Still, practical realization of the spectrometric (in fact, multi-monochromatic) approach in digital radiography is possible. This, however, would require substantial changes in the design of detector arrays, which should be matched to the emitter of the receiving-detecting circuit operating in the current mode of signal measurement. In the existing dual-energy scanners for customs inspection and medicine, the LED is always located before the HED (a sandwich design [1]) and is therefore the LED is sensitive to high energies. This leads to incomplete separation of the energies, which can be avoided using multi-monochromatic filters [14, 15]. They make it possible to single out narrow energy ranges of low, medium and high energies, separating at the same time the radiation beams of different energies (see Fig. 6), which improves the image quality of the inspected object even in the case when the receiver is not a detector, but an ordinary X-ray film.

## IV. CONCLUSIONS

Experimental and theoretical results obtained in the present work show that the use of the spectrometric approach can substantially improve the quality of identification of organic substances by their effective atomic number. The proposed procedure involves an array of values of parameter $P$, proportional to the degree of attenuation of the quanta flux in the low-energy range, obtained using special reference samples prepared in advance.

As it follows from the presented theory, in a more general case the role of parameter $P$ can be played by the ratio of separate integral signals $P = I_L / I_R$, recorded in low- or high-energy spectral range, respectively. A one-to-one correspondence should be expected between logarithms of this value and the effective atomic number. Therefore, the higher the spectrometric response accuracy (i.e., better energy resolution of the detector in the spectrometric mode), the better the accuracy of determination of the effective atomic number and other material parameters. This also leads to a conclusion that the use of monochromatic filters for separation of energies (and the corresponding signals) in multi-energy radiography is promising and expedient.

The proposed procedure can be used, e.g., for calibration of the color palette of observation monitors of customs inspection or medical diagnostics systems. This allows broadening of the range of substances that can be reliably identified, including light organic materials, e.g., explosives, drugs, etc.